\documentstyle[12pt]{article}
\date{}
\author{\small Gang Su and Bao-Heng Zhao\\
\small Department of Physics, Graduate School\\
\small Chinese Academy of Sciences, P. O. Box 3908\\
\small Beijing 100039, China}
\title{\large Long-Range Correlation of Electron Pairs
in the Hubbard Model at Finite Temperatures in Three Dimensions}

\begin{document}
\maketitle
\begin{abstract}
We show that in the translation invariant case and in the antiferromagnetic
phase, the reduced density matrix $\rho _2$ has no off-diagonal long-range
order of on-site electron pairs for the single-band Hubbard model on a cubic
lattice away from half filling at finite temperatures both for the positive
coupling and for the negative coupling. In these cases the model can not
give a mechanism for the superconductivity caused by the condensation of
on-site electron pairs and the nearest-neighbor electron pairs.

PACS numbers: 71.10.+x, 74.20.-z, 05.30.-d.
\end{abstract}

\newpage

It was proposed the possibility that the Hubbard model can give a mechanism
to characterize high temperature superconductivity\cite{1}. It has been
shown that the $\eta $ pairing states for the single band Hubbard model with
constant hopping matrix element for nearest neighbor sites have ODLRO\cite{2}%
. It has also been shown that in the ground state the model with negative
coupling has off-diagonal long-range order (ODLRO) on some bipartite lattice%
\cite{3}. However, as the temperature $T>0$ whether the reduced density
matrix $\rho _2$ has ODLRO is inconclusive. The purpose of this work is to
study the ODLRO of $\rho _2$ in the single-band Hubbard model with constant
hopping matrix element for nearest neighbor pairs on a cubic lattice at
finite temperatures. In the translation invariant case we obtain an equation
( (10) in the following) which is satisfied by the long-range correlation
functions of on-site electron pairs. From which we find that there is no
ODLRO of on-site electron pairs. Then we consider the case with an
antiferromagnetic background. In this case we will show that there is still
no ODLRO of on-site electron pairs.

The Hamiltonian of the model is
\begin{equation}
\label{1}H=-t\sum\limits_{\left\langle {\bf r,r}^{^{\prime }}\right\rangle
}(a{\bf _r^{\dagger }}a{\bf _{r^{^{\prime }}}}+b_{{\bf r}}^{\dagger }b_{{\bf %
r}^{^{\prime }}})+U\sum\limits_{{\bf r}}a{\bf _r^{\dagger }}a{\bf _r}b_{{\bf %
r}}^{\dagger }b{\bf _r}-\mu \sum\limits_{{\bf r}}(a{\bf _r^{\dagger }}a{\bf %
_r}+b{\bf _r^{\dagger }}b_{{\bf r}}),
\end{equation}
where $a_{{\bf r}}$ $(b{\bf _r})$ is the annihilation operator of the
electron with spin up (down) at site ${\bf r}.$ $U$ and $t$ are constant,
and $\mu $ is the chemical potential. $\left\langle {\bf r,r}^{^{\prime
}}\right\rangle $ is a pair of nearest-neighbor sites. The lattice is cubic
in the three dimensional space. We take the lattice spacing to be unity.

Define the $\eta -$operators as those in \cite{2,4,5},
\begin{equation}
\label{2}\eta _{-}=\sum\limits_{{\bf r}}e^{-i{\bf \pi \cdot r}}a{\bf _r}b%
{\bf _r},\,\ \eta _{+}=\eta _{-}^{\dagger },\,\ \eta _z=\frac 12\sum_{{\bf r}%
}(a_{{\bf r}}^{\dagger }a{\bf _r}+b_{{\bf r}}^{\dagger }b{\bf _r}-1),
\end{equation}
where{\bf \ $\pi $}$=(\pi ,\pi ,\pi ).$ They satisfy
\begin{equation}
\label{3}\left[ \eta _{+},\eta _{-}\right] =2\eta _z,\,\ \left[ \eta _{\pm
},\eta _z\right] =\mp \eta _{\pm },
\end{equation}
and
\begin{equation}
\label{4}\eta _{-}H=(H+U-2\mu )\eta _{-}.
\end{equation}

Consider the average over a grand canonical ensemble
\begin{equation}
\label{5}\left\langle \eta _{+}\eta _{-}\right\rangle =\frac 1ZTr(\eta
_{+}\eta _{-}e^{-\beta H}),
\end{equation}
where $Z$ is the grand partition function; $\beta $ the inverse temperature.
Using (4) and cyclically permuting the factors under the trace, we obtain $%
\left\langle \eta _{-}\eta _{+}\right\rangle =\left\langle \eta _{+}\eta
_{-}\right\rangle e^{\beta (U-2\mu )}.$ Then the first one of (3) gives,
when $U\neq 2\mu ,$%
\begin{equation}
\label{6}\left\langle \eta _{+}\eta _{-}\right\rangle =\frac{2\left\langle
\eta _z\right\rangle }{1-e^{\beta (U-2\mu )}}.
\end{equation}
{}From (2) we have $\left\langle \eta _z\right\rangle =(N-M)/2,$ where $%
N=\sum_{{\bf r}}\left\langle a_{{\bf r}}^{\dagger }a_{{\bf r}}+b_{{\bf r}%
}^{\dagger }b{\bf _r}\right\rangle $ is the average total number of
electrons, $M$ the total number of lattice sites. Denoting the density of
the number of electrons as $D=N/M,$ (6) leads to
\begin{equation}
\label{7}\frac 1M\left\langle \eta _{+}\eta _{-}\right\rangle =\frac{D-1}{%
1-e^{\beta (U-2\mu )}}.
\end{equation}
$U\neq 2\mu $ corresponds to the case away from half filling. We will only
consider this case below. As $T=\beta ^{-1}=finite$ and away from half
filling, the right hand side of (7) shows that $\frac 1M\left\langle \eta
_{+}\eta _{-}\right\rangle $ is a finite intensive quantity, which is still
finite in the thermodynamic limit ( $M\rightarrow \infty ,N\rightarrow
\infty $ with $D=N/M$ and $\beta $ fixed).

On the other hand, by the definition of $\eta _{\pm }$, we can write
\begin{equation}
\label{8}\frac 1M\left\langle \eta _{+}\eta _{-}\right\rangle =\frac
1M\sum\limits_{{\bf r,s}}\left\langle b{\bf _r^{\dagger }}a{\bf _r^{\dagger }%
}a_{{\bf s}}b_{{\bf s}}\right\rangle e^{i{\bf \pi \cdot (r-s})}.
\end{equation}
First we consider the case without the spontaneous breaking of the
translation symmetry on the lattice, and take the periodic boundary
condition. We have
\begin{equation}
\label{9}\frac 1M\left\langle \eta _{+}\eta _{-}\right\rangle =\sum\limits_{%
{\bf r}}\left\langle b{\bf _r^{\dagger }}a_{{\bf r}}^{\dagger
}a_ob_o\right\rangle e^{i{\bf \pi \cdot r}}.
\end{equation}
In the thermodynamic limit, the right hand side of (9) becomes an infinite
series. As we show above that $\frac 1M\left\langle \eta _{+}\eta
_{-}\right\rangle $ is finite, thus the infinite series $\sum\limits_{{\bf r}%
}\left\langle b{\bf _r^{\dagger }}a{\bf _r^{\dagger }}a_ob_o\right\rangle
e^{i{\bf \pi \cdot r}}$ is convergent .

Let us take the thermodynamic limit first, then, for the aim to investigate
the behavior of $\left\langle b{\bf _r^{\dagger }}a{\bf _r^{\dagger }}%
a_ob_o\right\rangle $ at large $\left| {\bf r}\right| ,$ consider a cube of
volume $(2S)^3$ in the infinitely large system. The center of the cube is at
the origin of the coordinates. Let $f(L)=\sum\limits_{faces}\left\langle b%
{\bf _r^{\dagger }}a{\bf _r^{\dagger }}a_ob_o\right\rangle e^{i{\bf \pi
\cdot r}},$ where{\bf \ }${\bf r}=(l,m,n),$ $-L\leq l,m,n\leq L,$ $l,m,n,L$
are integers, and $L\leq S$. $\sum\limits_{faces}$ is the sum over the six
faces of the cube with volume $(2L)^3,$ i.e.,%
$$
\sum\limits_{faces}=\sum\limits_{l,m=-L \\ (n=\pm L)}^L
+\sum\limits_{n,l=-L \\ (m=\pm L)}^L
+\sum\limits_{m,n=-L \\ (l=\pm L)}^L.
$$
It is clear that the sum of $\left\langle b_{{\bf r}}^{\dagger }a_{{\bf r}%
}^{\dagger }a_ob_o\right\rangle e^{i{\bf \pi \cdot r}}$ over the cube of
volume $(2S)^3$ can be written as $\sum\limits_{{\bf r}}\left\langle b_{{\bf %
r}}^{\dagger }a_{{\bf r}}^{\dagger }a_ob_o\right\rangle e^{i{\bf \pi \cdot r}%
}=\sum\limits_{L=0}^Sf(L).$ Since as $S\rightarrow \infty $ this series is
convergent, we have lim$_{L\rightarrow \infty }f(L)=0$, namely,
\begin{equation}
\label{10}\lim _{L\rightarrow \infty }\sum\limits_{faces}\left\langle b_{%
{\bf r}}^{\dagger }a_{{\bf r}}^{\dagger }a_ob_o\right\rangle e^{i{\bf \pi
\cdot r}}=0.
\end{equation}
Eq.(10) is a strong restriction on the long-range correlation functions. We
will show that it determines $\left\langle b{\bf _r^{\dagger }}a_{{\bf r}%
}^{\dagger }a_ob_o\right\rangle $ completely in the limit $\left| {\bf r}%
\right| $ $\rightarrow \infty .$

{}From the algebraic approach to the quantum statistical mechanics we know
that the correlation functions in the equilibrium states of a pure
thermodynamic phase have the spatial cluster properties , which is a
rigorous result\cite{6}. It enables us to write
\begin{equation}
\label{11}\left\langle b_{{\bf r}}^{\dagger }a{\bf _r^{\dagger }}%
a_ob_o\right\rangle \rightarrow \left\langle b_{{\bf r}}^{\dagger }a_{{\bf r}%
}^{\dagger }\right\rangle \left\langle a_ob_o\right\rangle ,\,\,as\,\,\left|
{\bf r}\right| \rightarrow \infty .
\end{equation}
In the translation invariant case $\left\langle a{\bf _r}b{\bf _r}%
\right\rangle =\left\langle a_ob_o\right\rangle $, thus $\left\langle b{\bf %
_r^{\dagger }}a{\bf _r^{\dagger }}a_ob_o\right\rangle \rightarrow \left|
\left\langle a_ob_o\right\rangle \right| ^2,$ as $\left| {\bf r}\right|
\rightarrow \infty .$ In (11) $\left\langle a_{{\bf r}}b{\bf _r}%
\right\rangle $ should be understood as Bogolubov's quasi-average, i.e., a $%
U(1)$ symmetry breaking term was added to the Hamiltonian, and after taking
the thermodynamic limit it has been sent to zero\cite{7}. $\left\langle a_{%
{\bf r}}b{\bf _r}\right\rangle $ does not vanish, if there is on-site
electron pair condensition, otherwise it vanishes. In the calculation of $%
\left\langle b{\bf _r^{\dagger }}a_{{\bf r}}^{\dagger }a_ob_o\right\rangle $
the symmetry breaking term is not necessary. Since $b_{{\bf r}}^{\dagger }a_{%
{\bf r}}^{\dagger }a_ob_o$ is an invariant under the global $U(1)$ gauge
transformation, its usual average ( in the Hamiltonian without the symmetry
breaking term) and the quasi-average are the same\cite{7}.

It is a well established empirical fact that in the second-order phase
transition the correlation length is divergent at the critical point, and it
is finite away from the critical point. The modern theories of critical
phenomena (the scaling hypothesis, the renormalization group approach, etc.)
are based on it\cite{8}. We assume that this empirical fact works in the
present problem, which is consistent with the superconducting transition
being a second-order phase transition. Define $G({\bf r})=\left\langle b{\bf %
_r^{\dagger }}a_{{\bf r}}^{\dagger }a_ob_o\right\rangle -\left| \left\langle
a_ob_o\right\rangle \right| ^2$ for arbitrary ${\bf r}.$ Suppose that the
system is in a noncritical thermodynamic state, then $G({\bf r})$ can not
decay slower than $O(\left| {\bf r}\right| ^{-3})$ as $\left| {\bf r}\right|
\rightarrow \infty ,$ otherwise the correlation length for $G({\bf r})$ will
diverge to infinity at a noncritical point. So $\lim _{L\rightarrow \infty }$
$\sum\limits_{faces}G({\bf r})e^{i{\bf \pi \cdot r}}=0.$ Hence (11) shows
that we can write
\begin{equation}
\label{12}\lim _{L\rightarrow \infty }\sum\limits_{faces}\left\langle b{\bf %
_r^{\dagger }}a{\bf _r^{\dagger }}a_ob_o\right\rangle e^{i{\bf \pi \cdot r}%
}=\left| \left\langle a_ob_o\right\rangle \right| ^2\lim _{L\rightarrow
\infty }\sum\limits_{faces}e^{i{\bf \pi \cdot r}}.
\end{equation}
It can be shown that
\begin{equation}
\label{13}\sum\limits_{faces}e^{i{\bf \pi \cdot r}}=(-1)^L2,\,\,\,for ~L \geq
1.
\end{equation}
Then (10) leads to $(-1)^L2\left| \left\langle a_ob_o\right\rangle \right|
^2\rightarrow 0.$ Therefore we obtain $\left\langle a_ob_o\right\rangle =0,$
and by translation invariance,
\begin{equation}
\label{14}\left\langle a_{{\bf r}}b_{{\bf r}}\right\rangle
=0,\,\,at\,\,any\,\,site\,\,{\bf r}.
\end{equation}
Furthermore (11) and (14) show that
\begin{equation}
\label{15}\left\langle b_{{\bf r}}^{\dagger }a{\bf _r^{\dagger }}%
a_ob_o\right\rangle \rightarrow 0,\,\,as\,\,\left| {\bf r}\right|
\rightarrow \infty .
\end{equation}
Eq.(14) shows that there is no condensation of on-site electron pairs, and
(15) shows that the reduced density matrix $\rho _2$ has no ODLRO for
on-site electron pairs. These two statements are equivalent.

Using (14) and taking the quasi-average of $\left[ H,a{\bf _r}b_{{\bf r}%
}\right] ,$ it can be shown \cite{9}
\begin{equation}
\label{16}\left\langle a{\bf _r}b{\bf _{r^{^{\prime }}}}\right\rangle =0,
\end{equation}
where ${\bf r,r}^{^{\prime }}$ are nearest neighbor sites. From the cluster
property of correlation functions, we can write $\left\langle b{\bf %
_r^{\dagger }}a{\bf _{r^{^{\prime }}}^{\dagger }}a{\bf _s}b{\bf %
_{s^{^{\prime }}}}\right\rangle \rightarrow 0,$ where $\left| {\bf %
r-r^{^{\prime }}}\right| =0\,\,$or$\,\,1,\,\,\left| {\bf s-s}^{^{\prime
}}\right| =finite,$ and $\left| {\bf r-s}\right| \rightarrow \infty .$

The above results are obtained for the translation invariant case. The
antiferromagnetic order breaks the translation symmetry. But we can show
that the existence of antiferromagnetic order, if any, is not in
contradiction with (14)-(16). In the antiferromagnetic phase there are two
kinds of translations, i.e., the displacement $\Delta {\bf r}=(\Delta
x,\Delta y,\Delta z),\,\Delta x+\Delta y+\Delta z=even\;\,integer,$ and $%
\Delta x+\Delta y+\Delta z=odd\,\,integer.$ The first one does not break the
translation symmetry, but the second one does. The second kind of
translations is equivalent to exchange the up spins and the down spins, so
it is easily seen that
\begin{equation}
\label{17}\left\langle a{\bf _r}b{\bf _r}\right\rangle =e^{i{\bf \pi \cdot r}%
}\left\langle a_ob_o\right\rangle ,\,\,\left\langle b{\bf _r^{\dagger }}a%
{\bf _r^{\dagger }}a{\bf _s}b{\bf _s}\right\rangle =\left\langle b_{{\bf r-s}%
}^{\dagger }a{\bf _{r-s}^{\dagger }}a_ob_o\right\rangle .
\end{equation}
Thus (9) and (10) are still valid, but (12) should replaced by
\begin{equation}
\label{18}
\begin{array}{c}
\lim _{L\rightarrow \infty }\sum\limits_{faces}\left\langle b_{
{\bf r}}^{\dagger }a_{{\bf r}}^{\dagger }a_ob_o\right\rangle e^{i{\bf \pi
\cdot r}}=\left| \left\langle a_ob_o\right\rangle \right| ^2\lim
_{L\rightarrow \infty }\sum\limits_{faces}1 \\ =\left| \left\langle
a_ob_o\right\rangle \right| ^2\lim _{L\rightarrow \infty }2(12L^2+1).
\end{array}
\end{equation}
So we can show that (14)-(16) still hold, i.e., there is no ODLRO for
on-site electron pairs.

In the derivation of (14)-(16), the sign of $U$ plays no role whatsoever, so
the above results apply to the cases both for $U>0$ and for $U<0.$

In summary, we have shown that for the translation invariant case or in the
antiferromagnetic phase the reduced density matrix $\rho _2$ has no ODLRO of
on-site electron pairs for the single band Hubbard model on a cubic lattice
away from half filling at finite temperatures both for $U>0$ and for $U<0$.
In these cases the model can not give a mechanism for the superconductivity
caused by the condensation of on-site electron pairs and nearest neighbor
electron pairs. To derive the above results we have assumed that in the case
away from the critical point the correlation length is finite , and no other
{\it ad hoc }assumption is needed. The above discussion can be easily
generalized to any dimensions. Our results are not incompatible with \cite{2}
and \cite{3}, since their results are not the ensemble average at $T>0.$

{\bf Acknowledgments}

The authors are grateful to B. Y. Hou, H. T. Nieh, G. S. Tian and M. Yu for
helpful discussions. This work is supported in part by the NSF of China.

\end{document}